\documentclass[reprint,amsmath,amssymb,aps,prl]{revtex4-2}

\usepackage{graphicx,color}
\usepackage{dcolumn}
\usepackage{bm}
\usepackage{easyReview} 
\begin{document}

\title{Predicting quasibound states of negative ions}
\author{M. S. Safronova$^{1,2}$, C. Cheung$^1$, M. G. Kozlov$^{3,4}$, S. E. Spielman$^5$, N. D. Gibson$^5$, and C. W. Walter$^5$}
\affiliation{$^1$Department of Physics and Astronomy, University of Delaware, Newark, Delaware 19716, USA\\
$^2$Joint Quantum Institute, National Institute of Standards and Technology and the University of Maryland, College Park, Maryland 20742, USA \\
$^3$Petersburg Nuclear Physics Institute of NRC ``Kurchatov Institute'', Gatchina, Leningrad District, 188300, Russia\\
$^4$St. Petersburg Electrotechnical University LETI, Prof. Popov Str. 5, St. Petersburg, 197376, Russia\\
$^5 $Department of Physics and Astronomy, Denison University, Granville, OH  43023 USA}%

\date{\today}

\begin{abstract}
We demonstrated the accurate prediction of a quasibound spectrum of a negative ion using a novel high-precision theoretical approach. We used La$^-$ as a test case due to a recent experiment that measured energies of  11 resonances in its photodetachment spectrum attributed to transitions to quasibound states [C. W. Walter \textit{et al.}, PRA, in press (2020);  	arXiv:2010.01122]. We identified all of the observed resonances, and predicted one more peak just outside the range of the prior experiment. Following the theoretical prediction, the peak was observed at  the predicted wavelength, validating the identification.
 The same approach is applicable to a wide range of negative ions. Moreover, theory advances reported in this work can be used for massive  generation of atomic transition properties  for neutrals and positive ions needed for a variety of applications.
\end{abstract}

\maketitle


\paragraph{Introduction.}
	Negative ions are important for both fundamental and practical reasons, such as medical applications \cite{Pegg2004,Leimbach2020}.  They are key constituents of terrestrial and space-based plasmas \cite{Millar2017}, and they play crucial roles in many chemical reactions, as highlighted, for example, in the very recent study of the astatine negative ion \cite{Leimbach2020}. Beams of short-lived radioactive nuclei are needed for frontier experimental research in nuclear
structure, reactions, and astrophysics, and negatively charged radioactive ion beams have unique
advantages and  can provide the
highest beam quality with continuously variable energies \cite{Liu2017}. Laser cooling of negative ions may allow a sympathetic cooling of antiprotons for the production of cold antihydrogen for tests of fundamental symmetries
\cite{Kellerbauer2006,Cerchiari2018}.
From a fundamental standpoint, since the extra electron in a negative ion is not bound by a net Coulomb force, their properties critically depend on electron-electron correlation and polarization, giving valuable opportunities to gain insight into these  important multibody interactions \cite{Andersen2004, Ivanov2004, Eiles2018}.   Therefore, negative ions serve as key test systems for state-of-the art atomic structure calculations.
	
	Excited states of negative ions, both bound and quasibound states known as resonances, pose even more challenges and opportunities for both theory and experiment \cite{Pegg2004, Andersen2004, Buckman1994}.  They are important in low energy electron scattering from atoms and molecules \cite{Buckman1999, Schippers2019, Szmytkowski2020}, electron attachment \cite{Schippers2019, Fabrikant2017}, chemical reactions \cite{Mikosch2010, Meyer2017}, and photoabsorption \cite{Pegg2004, Andersen2004, Ivanov2004, Kjeldsen2006, Schippers2020}.  Recent studies of negative ion excited states have focused on a diverse range of aspects, including the possibility of laser cooling negative ions  \cite{Kellerbauer2006,OMalley2010,Walter2014,Jordan2015,Yzombard2015, Cerchiari2018,Tang2019}, negative ion resonance spectroscopy using ultralong-range Rydberg molecules \cite{Engel2019}, and resonances in inner-shell photodetachment from small carbon molecular negative ions \cite{Perry-Sassmannshausen2020}.  Clearly, progress in theoretical calculations of negative ion excited states would be very valuable for a wide variety of both practical applications and fundamental insights.
	

	In this work, we demonstrate for the first time that a high-precision relativistic hybrid approach that combines the configuration interaction and the coupled-cluster methods (CI+all-order) \cite{SafKozJoh09,Cerchiari2018} can be used to accurately predict the energies of quasibound states of a negative ion.
This method was designed for low-lying bound states and generally bound state approaches cannot be used to compute properties associated with levels beyond the ionization (or in this case photodetachment) threshold for reasons described below, but we have developed successful ways to extend this technique to quasibound states of complex negative ions.

As a test case in the present study, we used the negative ion of lanthanum, La$^-$, which is one of the most intriguing of all atomic negative ions. Whereas most negative ions only have a single bound state configuration because of the shallow, short-range nature of their binding potentials \cite{Andersen2004}, La$^-$ possesses multiple bound states of opposite parity \cite{OMalley2010, Walter2014}.  Indeed, La$^-$ has the richest spectrum of bound-bound electric-dipole transitions yet observed for any atomic negative ion \cite{Walter2014}, and it has emerged as one of the most promising negative ions for laser-cooling applications \cite{OMalley2010, Pan2010, Walter2014, Jordan2015, Cerchiari2018}.  Beyond the complex bound structure of La$^-$, very recent measurements of its photodetachment spectrum have revealed a remarkably rich near-threshold spectrum including at least 11 prominent resonances due to excitation of quasibound negative ion states in the continuum \cite{Walter2020b}.  This recent observation of its photodetachment spectrum allowed for an immediate test of our theoretical predictions of the quasibound state structure of La$^-$. We were able to identify all of the 11 observed resonances (peaks), and predicted several peaks that were too weak to be observed in \cite{Walter2020b}. Our theoretical resonance energies agree with experiment to 0.03-0.3\% for ``narrow'' peaks and to 2.3-3\% for ``wide'' peaks associated with higher energy levels.   We also predicted that there should be a resonance peak just outside the photon energy range of the original experiment. Following our prediction, the peak was observed at exactly the predicted position, validating the identification; this new measurement is reported here.

 While we used La$^-$ as an example, the same approach is applicable to a wide range of  negative ions. Moreover, we developed a way to reliably extract hundreds of states in the framework of the CI+all-order method instead of just a few bound states.
 This advance will allow to significantly extend the applicability of the CI+all-order method for neutrals and positive ions, allowing for massive data generation (energies, transition rates, lifetimes, branching ratios, etc.) of a large part of the periodic table for a variety of applications.

  We start with a description of the theoretical computations and specific solutions that allowed us to extract the quasibound states and identify the measured resonances. Then, we describe a new experiment that found the peak predicted by the theory.

\paragraph{Theory.}
In 2018, the CI+all-order approach  was used to accurately predict energies of then unmeasured bound states of La$^-$ as well as transition rates and branching ratios relevant to the laser cooling of La$^-$ \cite{Cerchiari2018}.  In the CI+all-order method, the linearized coupled-cluster approach is used first to construct an effective Hamiltonian that includes core and core-valence correlations. Then, the many-electron wave function is obtained in the framework of the CI method as a
linear combination of all distinct many-electron states of a given
angular momentum $J$ and parity:
$ \Psi_{J} = \sum_{i} c_{i} \Phi_i.$
The energies and wave functions of the low-lying states are
determined by diagonalizing this effective Hamiltonian. La$^-$ is considered as a system with four valence electrons and Xe-like
54-electron core. The CI+all-order method uses Dirac-Hartree-Fock one-electron wave functions for the low-lying valence electrons; $6s$, $5d$, $4f$, $6p$, $7s$, and $7p$ in the present work. We use a finite basis method to construct all other orbitals (up to $35spdfghi$) in a
 spherical cavity using B-splines. Such an approach discretizes the continuum spectrum: a sum over the finite basis is equivalent (to a numerical precision) to the sum over all bound states and integration over the continuum. The obvious downside of this approach is
 the limitation of its applicability to relatively low-lying bound states. For example, even for the largest practical size of the cavity (a few hundreds atomic units) the method is limited to the orbitals with the principal quantum number less than 20, so higher Rydberg, or other delocalized states cannot be treated accurately.
 The situation for negative ions is more favourable, where there are (if any) only a few bound states, no usual Rydberg series, and quasibound states (if any) are still highly localized.
\begin{table} [t]
\caption{\label{tab1} Quasibound levels of La$^-$ energy levels in meV. All energies are counted from the $^3F_2$ even ground state.  Levels labelled $A$, $B$, $C$, and $D$ in experimental work \cite{Walter2020b} are assigned terms.
Calculated $g$-factors  are compared with the nonrelativistic values (NR) in the last two columns. }
\begin{ruledtabular}
\begin{tabular}{llccccc}
\multicolumn{1}{l}{Level}&\multicolumn{1}{c}{Term}& \multicolumn{1}{c}{Theory }& \multicolumn{1}{c}{Expt.}&  \multicolumn{1}{c}{Diff.(\%)} &\multicolumn{2}{c}{$g$-factor}\\
\multicolumn{5}{c}{} &\multicolumn{1}{c}{NR}&\multicolumn{1}{c}{CI+all}\\
\hline
$6s^25d6p      $&  $^3P_1 $&  567.0 &          &         &&              \\
$6s^25d6p      $&  $^3P_2 $&  643.2 &          &         &&                \\[0.3pc]
$6s5d^2(^4F)6p $&  $^5G_2 $& 725.0 &    723.34(4) &   -0.2\%   &0.333 &0.347 \\
$6s5d^2(^4F)6p $&  $^5G_3 $& 763.0 &    761.26(7) &   -0.2\%   &0.917 &0.924  \\
$6s5d^2(^4F)6p $&  $^5G_4 $& 814.1 &    811.27(4) &   -0.3\%   &1.150 &1.150  \\
$6s5d^2(^4F)6p $&  $^5G_5 $& 877.7 &          &            &1.267 &1.266  \\
$6s5d^2(^4F)6p $&  $^5G_6 $& 955.7 &          &            &1.333 &1.333  \\[0.3pc]
$6s5d^2(^4F)6p $&  $^5F_1 $& 900.4 &          &            &0.000 &0.083  \\
$6s5d^2(^4F)6p $&  $^5F_2 $& 920.1 &          &            &1.000 &1.001  \\
$6s5d^2(^4F)6p $&  $^5F_3 $& 953.3 &     979.3(11)&    2.7\% &1.250 &1.231     \\
$6s5d^2(^4F)6p $&  $^5F_4 $& 1005.9&          &            &1.350 &1.312    \\
$6s5d^2(^4F)6p $&  $^5F_5 $& 1068.0&          &            &1.400 &1.386    \\
\end{tabular}
\end{ruledtabular}
\end{table}

  There are two major problems in using the CI+all-order method to find quasibound states of negative ions.
  The first problem is the separation of true quasibound states from spurious ``continuum-like'' states that are artifacts of the finite basis (i.e. states containing orbitals with $n>9$ that do not fit inside the cavity and represent near-continuum states).  We solved this issue by running two set of calculations
   that were identical with the exception of the size of the cavity. We have theorized that the cavity size will affect the  number and energies of the spurious states.
  The bound and quasibound states will not be affected as the smaller cavity size is chosen to fit them inside the cavity (we expect quasibound states to be well localized). We find that our supposition is correct and the energies of the quasibound states are indeed stable with the change in the cavity size from 60 a.u. to 85 a.u. The difficulty of this approach comes from the second problem: a large number of spurious states drastically affect convergence of the iterative procedure used by the CI, which is already very poor for negative ions making the computations prohibitively long.  Furthermore, the convergence procedure was known to break down when some states reached convergence while other closely-lying states were still strongly varying. We separated the computation into seven different ones, each for a single value of the total angular momentum from $J=0$ to $J=6$ to improve convergence as well as resolved the issue of disparate convergence levels.

  Building upon the MPI version of the CI code developed in \cite{CheSafPor20}, we improved both efficiency and memory use, allowing to run such a large number of already complicated computations in a short time. Each of the computations contained 110\,000 - 186\,000 configurations, corresponding to 4-6.6 million Slater determinants and requiring at least 100 iterations (where usual is under 20).
 We computed a total of 74 odd states with $J=0-6$ and identified eight of these states as known bound states and twelve more states as quasibound states. We verified that the bound states agree with experiment to 0.1-2\%. We find that dominant configurations for ``spurious'' states contain a large fraction of the $np$ electrons with $n>8$ unlike the quasibound states where configurations with $6p$ and $7p$ dominate.

 The energies of quasibound  states are listed in Table~\ref{tab1} relative to the $^3F_2$ even ground state (detachment threshold 557.546(20) meV \cite{Lu2019,Blondel2020}).
Two of the quasibound  states complete the $^3P_J$ triplet, with $^3P_0$ state being the last bound state. We classify the remaining 10 states as 2 quintets, $6s5d^26p$  $^5G$ and  $^5F$. Both can be formed by attachment of a $6p$ electron to the $6s5d^2(^4F)$ excited states of La. To verify our term assignments we compute the $g$-factors for all states and compare them to the $g$-factors obtained from the non-relativistic Land\'{e} formula.
We find a near perfect agreement of the calculated $g$-factors with the non-relativistic values, see the last two columns of Table~\ref{tab1}, unambiguously confirming our term identification.

\begin{figure} [t]
\includegraphics[width=86mm]{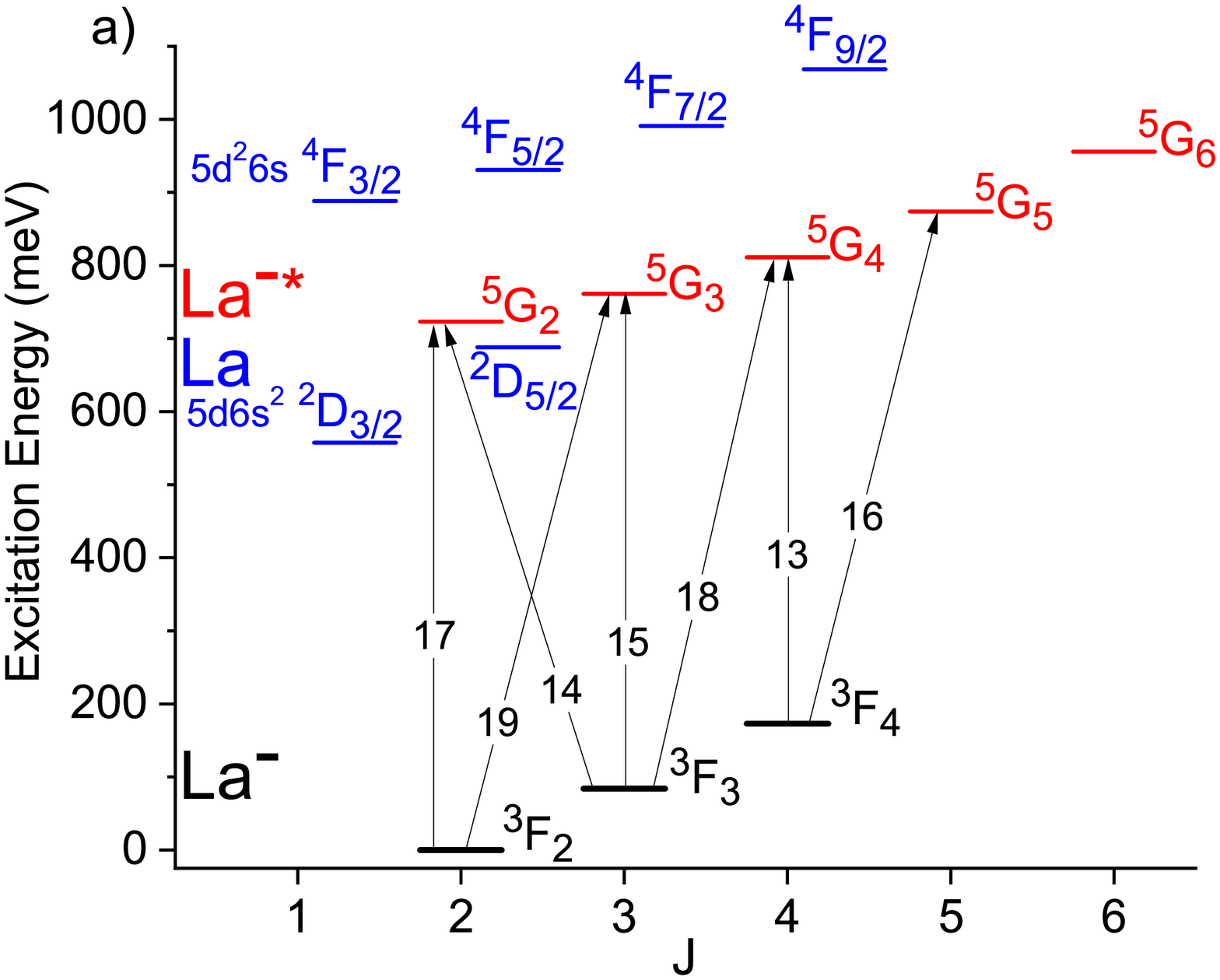}
\includegraphics[width=86mm]{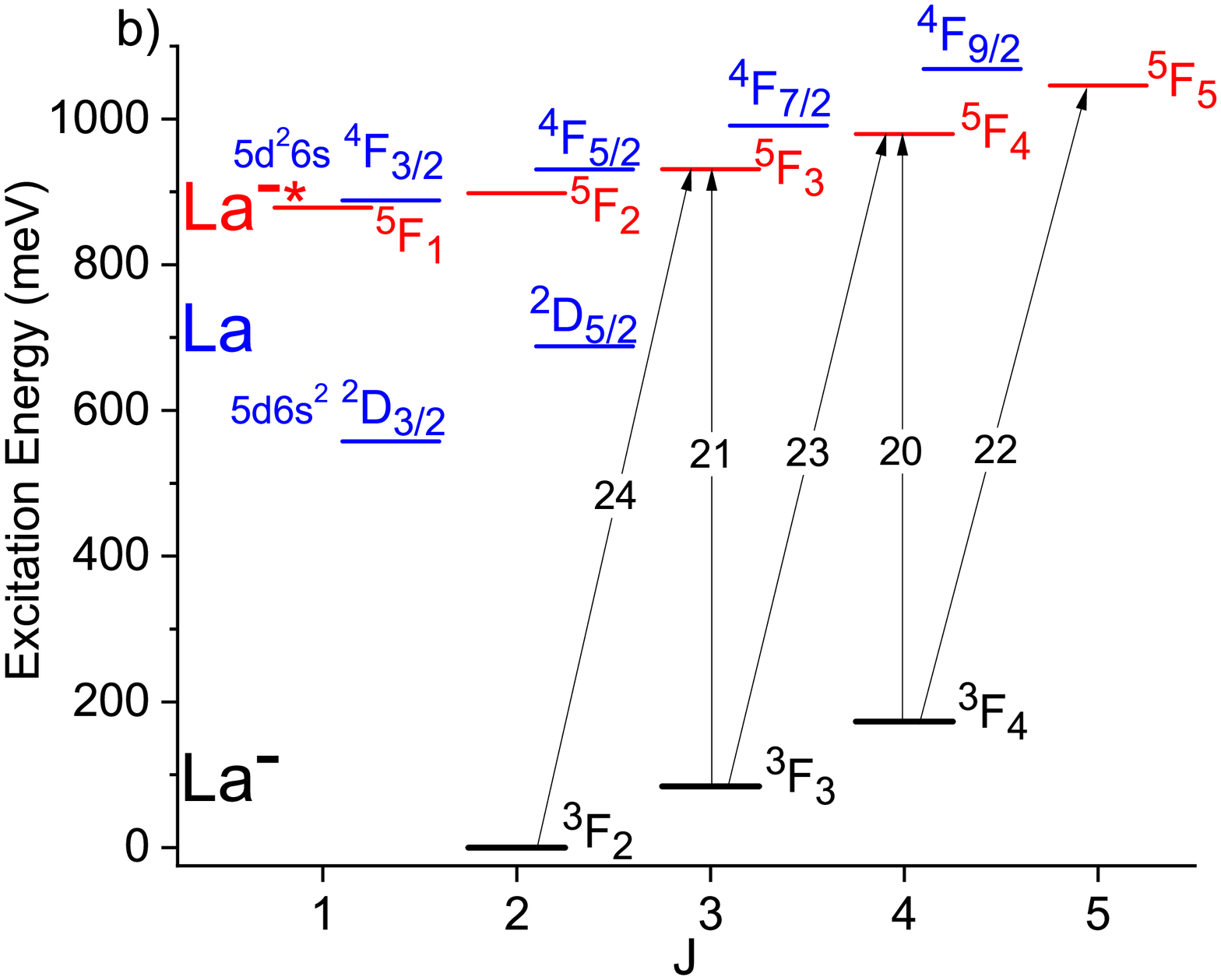}
\caption[]{\label{fig1}Partial energy level diagram of relevant bound states of La$^-$ (black), neutral La (blue), and  quasibound excited states of La$^-$ (red) in the a) $^5$\textit{G} and b) $^5$\textit{F} manifolds.  Numbered arrows indicate resonance transitions observed previously by our group \cite{Walter2020b} (Peaks 13-23) and in the present study (Peak 24) that have been identified in the present study.}
\end{figure}

The dipole selection rules allow for eight transitions from the three lowest-lying $6s^25d^2$ $^3F_{2,3,4}$ even states to the $^5G$ odd levels and nine
transitions to the $^5F$ odd levels. There are no allowed transitions involving the $^5G_6$ level. The transition energies for these 17 transitions are listed in Table~\ref{tab2}, together with the identification of peaks observed in \cite{Walter2020b} and their measured energies. All ``narrow'' (\textless 1 meV width) peaks 13-19 observed in \cite{Walter2020b} involve the $^5G$ levels. Due to excellent agreement of the theoretical predictions with the measured energies, all of these peaks  were straightforward to identify. All of the transition energies agree to 0.03-0.3\%. The only allowed transition that was not observed in \cite{Walter2020b} is $^3F_4 \rightarrow\,^5G_3$, which is expected  to be weaker than the observed $^3F_{2,3} \rightarrow\, ^5G_3$ transitions.
All observed transitions are  illustrated in Fig.~\ref{fig1} a) which shows a partial energy level diagram of relevant states of La$^-$ and La showing quasibound excited states in the  $^5$\textit{G} manifold.  The numbered arrows indicate resonance transitions that have been assigned in this study.
\begin{table} [t]
\caption{\label{tab2} Identification of peaks observed in \cite{Walter2020b}. Transition energies are given in meV. The recommended values  given in
``recomm.'' column are shifted by 22~meV from the \textit{ab initio} values. Last column gives the difference between the experimental and theoretical values in meV.}
\begin{ruledtabular}
\begin{tabular}{lccccc}
\multicolumn{1}{c}{Transition}&\multicolumn{2}{c}{Theory}&  \multicolumn{1}{c}{Expt.}
&\multicolumn{1}{c}{Peak}    &\multicolumn{1}{c}{Diff.} \\
\multicolumn{1}{c}{}&\multicolumn{1}{c}{\textit{ab initio}}&  \multicolumn{1}{c}{recomm.}
&\multicolumn{3}{c}{}   \\
\hline
$^3F_2\rightarrow\,^5G_2$&   725.0   &      &723.34(6)         &17  & 1.7 \\
$^3F_3\rightarrow\, ^5G_2$&   640.5   &      &639.41(5)         &14  & 1.1 \\
$^3F_2\rightarrow\, ^5G_3$&   763.0   &      &761.24(9)         &19  & 1.8  \\
$^3F_3\rightarrow\, ^5G_3$&   678.5   &      &677.36(5)         &15  & 1.1 \\
$^3F_4\rightarrow\, ^5G_3$&   587.5   &      &not observed          &    &      \\
$^3F_3\rightarrow\, ^5G_4$&   729.6   &      &727.32(3)         &18  & 2.3  \\
$^3F_4\rightarrow\, ^5G_4$&   638.6   &      &638.41(3)         &13  & 0.2  \\
$^3F_4\rightarrow\, ^5G_5$&   702.2   &      &701.01(4)         &16  & 1.2   \\ [0.5pc]
$^3F_2\rightarrow\, ^5F_1$&   900.4   &876.4 &  not observed        &    &      \\
$^3F_2\rightarrow\, ^5F_2$&   920.1   &898.1 &  blended with 23 &    &       \\
$^3F_3\rightarrow\, ^5F_2$&   835.6   &813.6 &  not observed       &    &         \\
$^3F_2\rightarrow\, ^5F_3$&   953.3   &931.3 &  predicted  &    &               \\
            &           &      &  observed  930.5(9)*          &    & \\
$^3F_3\rightarrow\, ^5F_3$&   868.8   &846.8 &  847.8(9)        &21  & 21.0    \\
$^3F_4\rightarrow\, ^5F_3$&   777.8   &755.8 &  not observed        &    &          \\
$^3F_3\rightarrow\, ^5F_4$&   921.4   &899.4 &  895.6(19)       &23  & 25.8      \\
$^3F_4\rightarrow\, ^5F_4$&   830.4   &808.4 &  806.3(13)       &20  & 24.1      \\
$^3F_4\rightarrow\, ^5F_5$&   892.5   &870.5 &  872.1(12)       &22  & 20.4       \\
\end{tabular}
\end{ruledtabular}
\flushleft{*Present work}
\end{table}

The remaining ``wide'' (\textgreater 1 meV width) peaks 20-23  in the spectrum are associated with transitions to the $^5F$ multiplet.
Peaks 20 and 23 have to involve the same $^5F_J$  level, as they are separated by 89~meV, which matches the energy difference of the $^3F_3$ and $^3F_4$ even states \cite{Walter2014, Lu2019, Blondel2020}. However, complete identification of the peaks 20-23 is more complicated as there are multiple ways to match these observed transitions to
theory predictions. We expect theory to be less accurate for these higher states due to stronger configuration mixing.
 The study of the fine-structure splittings within the $^5F$ quintet shows these to be regular, i.e. matching non-relativistic values to within a few meV. Therefore, we expect similar differences
between theory and experiment for all 4 measured transitions, with the deviation not exceeding  a few ($\sim$5) meV. This requirement leaves only  one possible identification of peaks 20-24 given in Table~\ref{tab2} in which all 4 measured energies  differ from the theory
by 20-25~meV. We predict that 3 transitions where total angular momentum $J$ is lower for the quasibound
state than for the even state  were too weak to be observed. Two of the transitions, $^3F_3\rightarrow\,^5F_4$ and $^3F_2 \rightarrow\,^5F_2$,
 have nearly the same energy resulting in blending of two transitions in Peak 23; note that the expected separation of these two transitions of only 1.3 meV is substantially less than Peak 23's width of 8.8(18) meV \cite{Walter2020b}).  To improve theory prediction for other peaks, we shift the calculated energies by 22~meV and list these recommended values in  the ``recomm.'' column, with expected uncertainties of less than 5 meV.

Importantly, from our identification of the quasibound state structure we expect a new resonance photodetachment peak associated with the  $^3F_2\rightarrow \,^5F_3$ transition at slightly higher energy than the previously measured spectrum of Walter \textit{}{et al.} \cite{Walter2020b}.
 Its predicted resonance energy can be calculated based on the energy of Peak 21, which is due to transition to the same $^5$\textit{F}$_3$ upper state but from a different lower state, $^3$\textit{F}$_3$.  Thus, the predicted energy of new Peak 24 is the energy of Peak 21 (847.8(9) meV) plus the La$^-$ ($^3$\textit{F}$_2$ - $^3$\textit{F}$_3$) fine structure splitting (83.941(20) meV \cite{Walter2014, Lu2019, Blondel2020}), yielding a predicted energy for Peak 24 of 931.7(9) meV.
Peak 20-23 identification and new Peak 24 are illustrated in Fig.~\ref{fig1} b), which shows transitions to the  $^5$\textit{F} manifold.

	\paragraph{Experiment.} To test the theoretical interpretation of the La$^-$ resonance spectrum,
our previous measurements \cite{Walter2020b} were extended to slightly higher photon energies to search for the predicted resonance due to the La$^-$ $^3$\textit{F}$_2$ $\rightarrow$ $^5$\textit{F}$_3$ transition near 931 meV.  The relative photodetachment cross section was measured as a function of photon energy using a crossed ion-beam-OPO laser-beam system described in detail previously \cite{Walter2010, Walter2011, Walter2020b}.  In the present study, the tuning range of the OPO was extended beyond its nominal short wavelength limit of 1350 nm by manually controlling its crystal in order to measure photodetachment from 920 - 948 meV (1350 - 1310 nm).

Figure 2 shows the La$^-$ photodetachment spectrum from Walter \textit{et al.} \cite{Walter2020b} together with the present measurements above 920 meV.
\begin{figure}
\includegraphics[width=86mm]{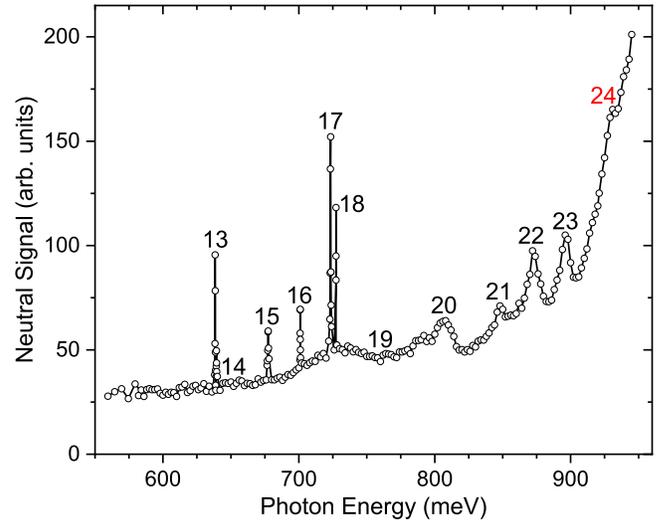}
\caption{\label{fig2}Measured La$^-$ photodetachment spectrum above the ground state threshold energy of 557.546 meV. Data below 920 meV are from our previous work \cite{Walter2014, Walter2020b}; data above 920 meV are from the present study.  The numbered peaks are due to resonant detachment via excitation of quasibound negative ion states; the newly observed Peak 24, which was predicted and verified in the present study, is indicated in red.}
\end{figure}
	The continuum photodetachment cross section rapidly rises above 920 meV in a nearly linear fashion due to the opening of photodetachment channels from bound states of La$^-$ to the La $^4$\textit{F} manifold.  Most importantly, the new measurements reveal an additional resonance peak, Peak 24, which appears as a weak hump in the cross section at an energy near 931 meV.  The measured photodetachment spectrum in the vicinity of the newly observed  Peak 24 is shown in Fig.~\ref{fig3}, together with a fit of the Fano resonance formula \cite{Fano1961} with a linear background continuum cross section.
\begin{figure}
\includegraphics[width=86mm]{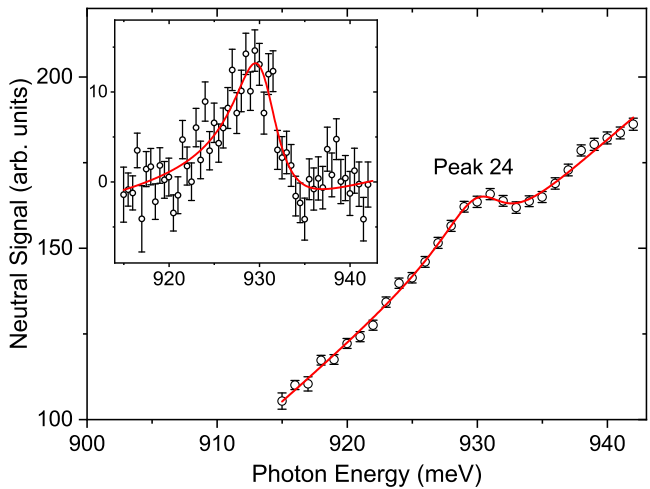}
\caption{\label{fig3}Measured photodetachment spectrum in the vicinity of the newly observed Peak 24, which is due to the La$^-$ $^3$\textit{F}$_2$ $\rightarrow$ $^5$\textit{F}$_3$ transition.  The solid line is a fit to the data of a Fano profile with a linear background.  The inset shows the remaining peak after the linear background has been subtracted from the measured neutral signal.}
\end{figure}
	The Fano profile provides an excellent fit to the data, yielding a resonance energy of 930.5(9) meV and peak width of 5.8(10) meV.

 The measured energy of Peak 24 of 930.5(9) meV agrees with the predicted value of 931.7(9) meV based on its theoretical identification as the $^3$\textit{F}$_2$ $\rightarrow$ $^5$\textit{F}$_3$ transition.  Furthermore, the measured widths of Peaks 21 and 24 (6.2(10) meV and 5.8(10) meV, respectively) are the same within uncertainties, as expected since the peak width depends on the lifetime of the $^5$\textit{F}$_3$ upper state which is the same for both peaks.  The agreement between the predicted and measured energy and width of the newly observed Peak 24 clearly verifies the present theoretical interpretation of the La$^-$ quasibound resonance spectrum and demonstrates the power of the calculational methods.

	It is important to note that the theoretical calculations were absolutely necessary to be able to find the new peak, since it is very weak ($<$ 8\% of the background signal) and situated on a steep slope due to a rapidly increasing continuum photodetachment cross section.
\paragraph{Conclusion.}
We demonstrated the ability to accurately predict the quasibound spectrum of negative ions. The accuracy of the theoretical calculations is unambiguously confirmed both by the identification of all resonance transitions in \cite{Walter2020b}, and, most importantly, prediction and observation of a new resonance peak. While we use La$^-$ as a test case, this method can predict quasibound states (if they exist) for other negative ions. The significant new computational advances in the present study that allowed efficient computation of a large number of energy levels are also applicable to neutrals and positive ions, allowing for generation of very large data sets for hundreds of levels needed for plasma physics, astrophysics, medical, and other  applications.

\begin{acknowledgments}
We thank Charles W.  Clark for insightful discussions.
This work was supported in part by U.S. NSF Grants No.\ PHY-1620687 and No.\ PHY-1707743. This research was performed in part under the sponsorship of the U.S. Department of Commerce, National Institute
of Standards and Technology.
 M.G.K. acknowledges support by the Russian
Science Foundation under Grant No.~19-12-00157.
The theoretical research was supported in part
through the use of Information Technologies resources at
the University of Delaware, specifically the high-performance Caviness computing cluster.
\end{acknowledgments}

%

\end{document}